\documentclass{ws-p9-75x6-50}
\def\Journal#1#2#3#4{{#1} {\bf #2}, #3 (#4)}

\def\SCI{{\em Science}}

\def\PRL{{\em Phys. Rev. Lett.}}
\def\PR{{\em Phys. Rev. }}
\def\Soviet{\em Zh. Eksp. Teor. Fiz.}

\def\PRB{{\em Phys. Rev.} B}
\def\PRC{{\em Phys. Rev.} C}
\def\PR{{\em Phys. Rev.}}
\def\PHYS{{\em Phys. Rep.}}
\def\MOL{{\em Mol. Phys. }}
\def\PRO{{\em Prog. Theor. Phys.}}
\def\CHEM{{\em J. Chem. Phys.}}
\def\LOW{{\em J. Low Temp. Phys.}}

\def\be{\begin{equation}}
\def\ee{\end{equation}}
\def\bea{\begin{eqnarray}}
\def\eea{\end{eqnarray}}

\begin{document}
\title{A MICROSCOPIC LOOK AT LIQUID HELIUM: THE $^3$He IMPURITY CASE}

\author{A. POLLS}

\address{
Departament d'Estructura i Constituents de la Mat\`eria, 
Universitat de Barcelona,\\
 E-08028 Barcelona, Spain\\E-mail:artur@ecm.ub.es}

\author{A. FABROCINI}

\address{
 Istituto Nazionale di Fisica Nucleare, 
Dipartimento di Fisica, Universit\`a di Pisa,\\ 
I-56100 Pisa, Italy \\E-mail:fabro@galileo.pi.infn.it}


\maketitle\abstracts{ 
The description of the properties of liquid Helium is a challenge for 
any microscopic many--body theory. In this context, we study the ground 
state and the excitation spectrum of one $^3$He impurity in liquid $^4$He 
 at $T=0$ with the aim of illustrating the power of the correlated basis
function formalism in describing heavily correlated systems. The strong 
interatomic interaction and the large density require 
the theory to be pushed to a high degree of sophistication. A many--body 
correlation operator containing explicit two-- and thre--particle 
correlation functions is needed to obtain a realistic ground state 
wave function, whereas a perturbative expansion including up to two phonon 
correlated states must be enforced to study the impurity excitation 
energies. The theory describes accurately the experimental spectrum 
along all the available momentum range. As empirically shown by the 
experiments, a  marked deviation from the quadratic Landau-Pomeranchuck 
behavior is found and the momentum dependent effective mass of the 
impurity increases of $\sim50~\%$ at $q\sim1.7~\AA^{-1}$ with respect to 
its $q=0$ value.  Although the main emphasis is given to the Correlated 
Basis Function theory, we present also comparisons with other 
methods, as diffusion Monte Carlo, variational Monte Carlo with shadow 
wave functions and time dependent correlations.}

\section{Introduction}

Atomic Helium fluids are an endless source of physical motivations
for both theorists and experimentalists. We have  so far 
accumulated a huge amount of experimental information and the 
activity is continuously pushed further to explore new situations. 
An example is the physical realization of almost two--
\cite{booktwo} (films) and one--dimensional \cite{hallock99}
 (nanotubes) systems and the consequent possibility to study the 
dependence on dimensionality of correlation effects. Helium clusters 
are another extremely promising field and 
the present experimental capabilities make possible to investigate 
the minimum number of atoms needed to have superfluidity 
\cite{vilesov99}.
For theoreticians, Helium liquids can be considered
as excellent laboratories to test many-body theories. In fact, 
the interaction is simple and depends only on the distance between 
the atoms.  In addition, the effects of  having different kinds
of quantum statistics may be studied, since here the quantum
behaviour has macroscopic manifestations. Actually, the fact that Helium 
remains liquid at zero temperature is a consequence of the large zero
point motion of the atoms in the fluid and can be considered as a 
macroscopic quantum effect.

 In spite of the enormous progress in the last two decades \cite{book1}
we have not yet achieved a  completely satisfying explanation 
of some  experimental facts. Let's just mention 
that a full microscopic description of superfluidity in liquid $^3$He is 
still missing. It is not possible, in a single presentation, to 
summarize the present status of the whole Helium physics field, so 
we will exploit the fact that in the last two years several microscopic 
many-body theories have been used to study the ground and excited states 
of one $^3$He impurity in liquid $^4$He at zero temperature and  use
this system to illustrate the power and   peculiarities
of these  techniques. The main emphasis will be
given to Correlated Basis Function \cite{fabro86,fabro98}
but  comparisons with the
results obtained with Shadow Wave Function\cite{reatto99} (SWF),
 Time Dependent Correlations
\cite{kro98}(TDC) or Difusion Monte Carlo\cite{boro99}(DMC)
 will be given.
 
The  theoretical study of one $^3$He impurity in atomic liquid
 $^4$He  is very helpful in understanding  the properties of 
$^3$He-$^4$He mixtures. $^3$He and $^4$He are isotopes which 
follow Fermi and Bose statistics, respectively. They
can form a liquid mixture which remains stable at zero temperature, with a
maximum solubility of the $^3$He component at zero pressure of $\sim$6.6$\%$.
 Both types of statistics coexist in the mixture with the consequence 
 that the excitation spectrum is particularly rich. There are two types 
of excitations, whose lowest energy one corresponds to  
a particle-hole band associated to $^3$He quasi-particles. 
The quasi-particles have a single particle spectrum characterized by 
an effective mass, $m_3^*$, mainly due to the interaction with the $^4$He 
atoms.  A little higher in energy lies the collective phonon-roton branch,
 associated to $^4$He and little affected by the presence of the $^3$He 
component.  There is a marginal low momentum  admixture between the two 
branches  and they have been clearly separated in  recent 
inelastic neutron scattering experiments at low momentum transfer\cite{Fak90}.

As the $^3$He concentration, $x_3$, in the mixture is small, it is justified to
 study the $x_3\rightarrow$0 limit, or the impurity problem.
The analysis of the impurity behavior not only provides 
 clues to understand the mixture itself but also gives useful information
on pure $^4$He liquid. As we will see,  the impurity can be used
as a theoretical probe to examine the 
kinetic and potential energies of the bulk \cite{boronat89} system. 

The bulk  properties of helium liquid are well measured \cite{wilks67}, and  
a complete 
thermodynamic information is also available for $^3$He-$^4$He mixtures \cite{ebner70}.
 In particular, the chemical potential of the $^3$He impurity 
  at $^4$He saturation density ($\rho_0(^4He)=
0.02186 \AA^{-3}$) and zero temperature is $\mu_3 = -2.78$ K,  
to be compared with the chemical potential of pure $^4$He , 
$\mu_4= -7.17$ K, and with the one of liquid $^3$He, 
$\mu_3 = -2.5$ K  
at its own saturation density, $\rho_0(^3 He)=0.0163 \AA^{-3}$.

The recent inelastic neutron scattering experiments \cite{Fak90} 
at low momenta and low concentrations  have given access to 
 the impurity excitation spectrum. The experiments show a
 significant deviation  from the quadratic
 Landau-Pomeranchuck (LP) spectrum \cite{Landau48}, 
$\epsilon_{LP}(q)=\hbar^2 q^2/2m_3^*$. The measured 
spectrum is well described in a modified LP (MLP) form
\begin{equation}
\epsilon_{MLP}(q)=\frac {\hbar^2 q^2}{2 m_3^*}\,
\frac{1}{ 1\,+\, \gamma\ q^2}\, .
\label{eq:LPM}
\end {equation}
 Although there are still some uncertainties in the analysis of the 
data \cite{Fak90}, the estimated values of the MLP parameters,  
at $P=1.6$ bar and $x_3\sim0.05$ are 
 $m_3^*\sim 2.2~m_3$ and $\gamma\sim 0.13~\AA^2$. 
A recent analysis \cite{kro198} of  specific heat measurements 
\cite{yoro93} gives a slightly different zero momentum 
 effective mass of $m_3^* = 2.18~m_3$, taking into account 
the corrections due to the $^3$He finite concentration in the mixture. 

 The purpose of this contribution is to report the recent progress in 
the microscopic description of the impurity system. This will give us 
the opportunity to comment also on the present status of the description 
of the $^4$He liquid and enlight some aspects of the dynamics of 
$^3$He-$^4$He mixtures. 

\section{Static  properties}

For a quantum microscopic description we start from an empirical
 Hamiltonian, defined in terms of the masses of the atoms and of 
their mutual interactions. In the case of one impurity, the 
Hamiltonian reads
\begin{equation}
H=- \sum_{i=1}^{N_4} \frac {\hbar^2}{2 m_4}\nabla_4^2+ \sum_{i,j}^{N_4}
V(\mid 
{\bf r}_i^{(4)}-{\bf r}_j^{(4)} \mid) 
- \frac {\hbar^2}{2 m_3}\nabla_3^2 +\sum_{i=1}^{N_4}
 V(\mid {\bf r}^{(3)}-{\bf r}_i^{(4)}\mid .
\label{eq:hamil}
\end{equation}
Due to the isotopic character of the mixture, the interaction
is the same between the different pairs of atoms. 
A simple representation of the potential is the 
Lennard-Jones interaction
\begin{equation}
V(r)= 4 \epsilon \left [ \left( \frac {\sigma}{r} \right )^{12} - \left 
( \frac {\sigma}{r} \right )^{6} \right ],
\end{equation}
where $\epsilon =10.22$ K gives the depth of the potential and 
$\sigma =2.556$~ \AA~ defines the length scale. Nowadays, the more 
accurate Aziz potential \cite{Aziz} and its revised version 
\cite{Aziz1} HFD-B(HE) are used in realistic calculations. 
The He--He potential is characterized by a strong short range 
repulsion (such that the atoms, at a first approximation, can be 
considered as hard--spheres of diameter $\sim$2.6 \AA ) 
and a weak attraction at medium and large distances. 
An important feature to keep in mind is that the ionization energy of the 
atoms and the first excitation energy are large ( of the order of the eV) 
when compared with the energies which play a role in the physics of 
Helium liquids and that are of the order of the  Kelvin 
( 1 eV $\sim$ 11000 K). Therefore, the Helium atoms are truly the 
elementary constituents of the system.
 
The chemical potential of the $^3$He impurity is defined by 
\begin{equation}
\mu_3 = \frac {\langle \Psi(3;N_4) \mid H(3;N_4) 
\mid \Psi(3;N_4) \rangle}{\langle \Psi(3;N_4)\mid \Psi(3;N_4) \rangle }
   - 
\frac {\langle \Psi(N_4) \mid H(N_4) \mid \Psi(N_4) \rangle}{\langle \Psi(N_4) \mid
\Psi(N_4) \rangle },
\label{eq:chemical}
\end{equation}
i.e. as the energy difference  when one $^3$He atom is added to 
$N_4$ $^4$He atoms at constant volume.
 The quantities to be subtracted are of order $N_4$ 
while the result of order of unity.
 Therefore, one must explicitly take
into account these cancellations between large  quantities 
to have a good estimate of $\mu_3$ .

In a variational approach the next step consists in choosing 
 a suitable trial function ($\Psi_0=\Psi_0(3;N_4)$) for 
the ground state  of $N_4$ $^4$He 
atoms plus one $^3$He in a volume $\Omega$, 
taken in the $N_4,~\Omega \rightarrow\infty$ limit, at constant 
$^4$He density, $\rho_4=N_4/\Omega$. The background wave function 
is simply obtained by omitting the impurity in the wave function. 
The extended Jastrow--Feenberg correlated wave function
 \cite{feenberg},
\begin{equation}
\Psi_0(3;N_4)\,= F_2(3;N_4) \, F_3(3;N_4) \, ,
\label{eq:Psi_0}
\end{equation}
represents a realistic choice for $\Psi_0(3;N_4)$. 
$F_{2,3}$ are N--body correlation 
operators including explicit two-- and  three--body dynamical
correlations. $F_2$ is written as a product of two  
body Jastrow, $^3$He-$^4$He and $^4$He-$^4$He correlation functions, 
\begin{equation}
F_2(3;N_4)=\prod_{i=1,N_4} f^{(3,4)}(r_{3i}) \, 
  \prod_{m>l=1,N_4} f^{(4,4)}(r_{lm}) \, , 
\label{eq:F_2}
\end{equation}
and $F_3$ is given by a corresponding product of triplet correlations, 
$f^{(\alpha, \beta, \gamma)}({\bf r}_\alpha,{\bf r}_\beta,{\bf r}_\gamma)$.

The minimization  of the ground state energy gives, in principle,  
the correlation functions via the solution of the Euler equations,  
$\delta E_0/\delta f^{(\alpha, \beta)}=0$ and 
$\delta E_0/\delta f^{(\alpha, \beta ,\gamma)}=0$.
Most of the results presented in this paper have been obtained by 
using an analytical form for the three--body correlation and 
solving the Euler equations for the Jastrow factor within the 
Hypernetted Chain (HNC) integral equations technique for the 
distribution functions \cite{fabrocini84}. 

The parametrized triplet correlations have the Feynman form,
\begin{equation}
f^{(\alpha, \beta, \gamma)}
({\bf r}_\alpha,{\bf r}_\beta,{\bf r}_\gamma)
= \exp { \left [ - \frac{1}{2} \sum_{cyc} 
\xi(r_{\alpha \beta}) \xi(r_{\alpha \gamma})\hat r_{\alpha \beta}
 \cdot \hat r_{\alpha \gamma} \right ]}
\end{equation}
$\xi(r)$ being a parametrized variational function \cite{schmidt,usmani}.
An  optimization of the triplet correlations in a larger functional
space shows that the Feynman form is nearly optimal \cite{moroni95}.
Optimal three body correlations for the impurity in the bulk, solving 
the Euler equation for $f^{(\alpha, \beta, \gamma)}$, 
 have been recently obtained and used
 \cite{saarela93,saarel193}.  

For the sake of illustration we give the energy equations with 
two body correlations only. The 
cancellations occuring in the calculation of $\mu_3$ may be 
effectively dealt with by writing the expectation value
of $H(3;N_4)$ as $E_0^v(3;N_4) = E_4^v + \mu_3^v$, 
where $E_4^v$ is the energy of the medium (proportional to $N_4$) and 
$\mu_3^v$ is the chemical potential of the $^3$He atom (of the order of 
unity). In fact, the background energy, \ $E_4^v$, cancels  
in the difference (\ref{eq:chemical}). 
The background energy, obtained within the Jackson-Feenberg
prescription for the kinetic energy, is given by 
\begin{equation}
\frac {E_4^v}{N_4} = \frac {\rho_4}{2} \int d{\bf r} g^{(4,4)}(r) 
\left [V(r) - \frac {\hbar^2}{2 m_4} \nabla^2 \ln {f^{(4,4)}(r)} \right ].
\end{equation}
$\mu_3^v= \mu_{int} + \mu_{rea}$ is decomposed into the interaction
term $\mu_{int}$, 
\begin{equation}
\mu_{int}= \rho_4 \int d {\bf r} g^{(3,4)}(r) \left [ V(r) -
 \frac {\hbar^2}{4 m_{red}} \nabla^2 \ln {f^{(3,4)}(r)} \right],
\end{equation}
$m_{red}$ being the reduced mass $m_{red}= m_3 m_4/(m_3+m_4)$,
and $\mu_{rea}$, due to the medium rearrangement, 
\begin{equation}
\mu_{rea} = \frac {1}{2} \rho_4 \int d{\bf r} g_{rea}^{(4,4)}(r) \left [
V(r) - \frac {\hbar^2}{2 m_4} \nabla^2 \ln {f^{(4,4)}(r)} \right ].
\end{equation}
The distribution functions are defined by: 
\begin{equation}
g^{(4,4)}(r_{12})~+~\frac {1}{\Omega} g_{rea}^{(4,4)}(r_{12}) = 
\frac {N_4 (N_4-1)}{\rho_4^2}
\frac {\int d{\bf r}_3 d{\bf r_3} .... \mid \Psi(3;N_4)\mid^2}{\int d\Omega
\mid \Psi(3;N_4)\mid ^2},
\label{eq:distri1}
\end{equation}
\begin{equation}
g^{(3,4)}(r_{31})= \frac {\Omega N_4}{\rho_4} \frac {\int d{\bf r}_2 d{\bf r}_3 
...  \mid \Psi(3;N_4)\mid^2}{\int d \Omega \mid \Psi(3;N_4) \mid^2}.
\label{eq:distri2}
\end{equation}

One can also introduce the corresponding static structure functions, 
via the Fourier transforms:
\begin{equation}
S(k)\equiv S^{(4,4)}(k)=1~+~\rho_4 \int d{\bf r} e^{i {\bf k}\cdot {\bf r}}
 \left [ g^{(4,4)}(r)-1
\right ].
\label{eq:static}
\end{equation}
and
\begin{equation}
S^{(3)}(k)\equiv S^{(3,4)}(k)~=~ 
\rho_4 \int d{\bf r} e^{i {\bf k}\cdot {\bf r}} \left [ g^{(3,4)}(r)-1 
\right ].
\label{eq:staticI}
\end{equation}
In the $k \rightarrow 0$ limit
\begin{equation}
  \lim_{k\rightarrow 0} S^{(3)}(k) =-(1+\alpha)
\label{eq:limit}
\end{equation}
where $\alpha$ is the excess volume parameter, i.e. the relative increment
in the molar volume when the impurity is added to the system, keeping the
pressure constant. The mobility of the lighter mass $^3$He is 
larger and its  presence tends to increase the pressure. 
As a consequence, the volume must be enlarged  to keep the pressure constant. 
The experimental value of $\alpha$ at the $^4$He saturation density is 0.284.

The problem of calculating the expectation values has so been translated
into that of the evaluation of the distribution functions. Cluster 
expansion and integral equations provide a viable and effective 
tool to perform this task. To this aim, one  introduces the function 
$h= f^2 -1$  and expands the distribution functions in powers of $h$. 
The terms of the expansions are diagrammatically classified and summed 
up to infinite orders by Hypernetted Chain integral equations, whose 
properties have been pedagogically reviewed by Fabrocini and Fantoni, 
for both Bose and Fermi systems \cite{book1,fabro199}.  

The impurity problem has been often analyzed within the 
 Average Correlation Approximation (ACA) \cite{polls93}. 
ACA is obtained by taking the same dynamical correlation functions 
for the impurity and for the medium.  In this case,
$g^{(3,4)}=g^{(4,4)}$ and 
\begin{equation}
g_{rea}^{(4,4)}(r_{12})= \frac {\partial g^{(4,4)}(r_{12})}{\partial \rho_4},
\end{equation}
therefore
\begin{equation}
\mu_{int}^{ACA}= 
2 e(\rho_4) + \left ( \frac {m_4}{m_3} -1 \right ) t^{(4)}(\rho_4)
\end{equation}
where $e(\rho_4)$ and $t^{(4)}(\rho_4)$ are the  total and the 
kinetic $^4$He energies per particle, respectively.
$\mu_{rea}$ is then given in ACA by
\begin{equation}
\mu_{rea}^{ACA} = \frac {P(\rho_4)}{\rho_4} - e(\rho_4)
\end{equation}
$P(\rho_4)$ being the thermodynamical pressure. 
In this way the chemical potential is expressed as
\begin{equation}
\mu_3^{ACA} = 
e(\rho_4)+ 
\frac {P(\rho_4)}{\rho_4} + \left (\frac {m_4}{m_3} -1 \right )
 t^{(4)}(\rho_4) =
\mu_4 (\rho_4) + \left ( \frac {m_4}{m_3} -1 \right ) t^{(4)}(\rho_4).
\end{equation}
The chemical potential of the impurity in ACA is the chemical 
potential of pure
$^4$He corrected by a kinetic energy term properly scaled to take into 
account the different mass of the impurity.
 Actually this phormula can be easily generalized to 
 wave functions containing also n-body correlations 
\cite{baym66,polls93}. 
 A recent difusion Monte Carlo calculation\cite{boro99}, 
which contains optimized n-body correlations,  has given 
$t_4$= 14.32 K and $\mu_4 =-7.27$ K at saturation density.  
Employing these values, one gets $\mu_3^{ACA}=-2.58$ K. 
If  the optimization of the correlations affecting the impurity is 
performed, the chemical potential value is closer to the experimental one.
       
An alternative trial wave function for Bose systems has been recently 
proposed: the so called Shadow wave function method\cite{vitiello}. 
It has been succesfully used for microscopic calculations on liquid 
and solid $^4$He at zero temperature \cite{moroni}
and has been  generalized to the impurity system\cite{reatto99}. In 
the SWF approach the wave function is given by
\begin{equation}
\Psi_{SWF}(R) = \int dS~ F_{SWF}(R,S)
\end{equation}
where $R= \left (\vec r_3,\vec r_1, ..., \vec r_N \right)$ 
are the coordinates of
the particles and $S=\left (\vec s_3, \vec s_1,... \vec s_N \right)$ is a set 
of auxiliary variables, representing centers of motion of the particles and 
integrated over the whole space. The interparticle correlations are
embedded in
\begin{equation}
F_{SWF}= \varphi_p(R) \times f^3_{ps}(\mid \vec r_3 -\vec s_3\mid) \times \prod_{i=1}
^{N_4} f_{ps}(\mid \vec r_i -\vec s_i \mid) \times \varphi_s(S).
\end{equation}
$\varphi_p(R) = \prod_{i=1}^{N_4} f_p^3(\mid \vec r_i - \vec r_3 \mid ) \times 
\prod _{i < j} f_p(\mid \vec r_i - \vec r_j \mid)$ is a Jastrow factor, 
as well as 
$\varphi_s(S) = \prod_{i=1}^{N_4} f_s^3(\mid \vec s_i - \vec s_3 \mid )
\times \prod_{i<j} f_s(\mid \vec s_i - \vec s_j \mid )$. 
$\Psi_{SWF}$ includes
correlations at all orders (pair, triplet, ...) 
through the shadow correlations,
after integrating over $S$. One of the main features of $\Psi_{SWF}$ is the
possibility to describe both liquid and crystalline phases within the same
functional form, respecting Bose symmetry and translational invariance.  
The expectation values between these trial wave functions are usually
calculated by Monte Carlo techniques. 
The extension of ACA to the shadow variables approach, i.e. 
by taking the correlation factors $f_p, f_s$ and $f_{ps}$
as recently optimized for pure $^4$He \cite{moroni} also for 
the impurity correlations, gives $\mu_3=-2.43$ K at saturation density,
 close  to the ACA value obtained with the DMC results as input. 
In the case of the SWF the three-, four- and n-body correlations are generated 
through the correlations with the shadow variables, while in the correlated 
basis theory they are explicitely introduced  
in the variational wave function and in the DMC method they 
 are incorporated and/or optimized in the DMC algorithm    
 for solving the many--body Schr\"odinger equation.

The density dependence of the pure $^4$He energy and 
of the $^3$He impurity chemical potential is reasonably well 
described by CBF if the correlations are properly optimized and the 
cluster expansion diagrams are fully summed, with the inclusion 
of the {\em elementary} ones. CBF shows a good agreement with the results 
obtained within the two other approaches briefly discussed above. 
Even if the finite size of the simulation  box introduces some uncertainties
in the Monte Carlo methods and  some limitations in the study the long 
range behavior of the distribution functions, DMC provides 
in principle the most accurate results. In fact, by 
 using the HFD-B(HE) Aziz potential \cite{Aziz1} it 
 reproduces the $^4$He experimental saturation density $\rho_0$, 
as already mentioned it gives $\mu_4^{(DMC)}(\rho_0)=-7.27(1)$ K and 
provides $\mu_3^{(DMC)}(\rho_0)=-2.79(25)$ K and $\alpha=0.284(10)$. 
The kinetic and potential energies of the pure liquid $^4$He at $\rho_0$ 
 are $t(\rho_0)=14.32(5)$ K and $v(\rho_0)=-21.59(5)$ K, respectively. 

Before leaving this subsection, we show in Figure 1 the static-structure
functions $S(k)$ and $S^{(3)}(k)$. They are the main inputs 
in the calculation of the excitation spectrum and have been obtained at 
 the experimental saturation density with the Aziz potential and solving 
the Euler-Lagrange equations for the two-body correlation functions, 
keeping the triplet correlations fixed \cite{fabrocini84}.  
$S^{(3)}(k)$ has the proper behavior at $k \rightarrow 0$  
(\ref{eq:limit}). 
The ACA cannot reproduce this limit because $S^{(3)}_{ACA}(k)=S(k)-1$. 
 
\begin{figure}
\begin{center}
\epsfysize=2.5in \epsfbox{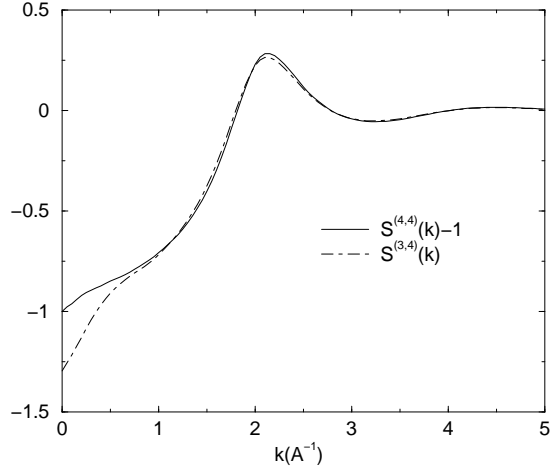}
\end{center}
\caption{ Pure $^4$He and impurity 
static structure functions at the $^4$He saturation density.}
\end{figure} 

\subsection {The $^3$He impurity as a probe in liquid $^4$He}

The difference of expectations values in Eq.(\ref{eq:chemical}) shows 
that $\mu_3$, as given by the equation, is not necessarily un upper bound 
to the true chemical potential.  However, if we assume that the trial wave 
function is the exact wave function of liquid $^4$He, 
then $\mu_3^{ACA}$ provides an exact upper bound to $\mu_3^{expt}$,
\begin{equation}
\mu_3^{expt}(\rho_4) \leq \mu_4^{expt}(\rho_4)+ 
\left ( \frac {m_4}{m_3}-1 \right ) t(\rho_4).
\end{equation}
Since $\mu_4^{expt}(\rho_4)$ and $\mu_3^{expt}(\rho_4)$ are experimentally 
known, the previous inequality can be used to establish a model independent 
lower bound to the kinetic energy of liquid $^4$He,
\begin{equation}
t(\rho_4) \geq t^{LB}(\rho_4) \equiv 
\left [\frac {m_3}{m_4-m_3} \right ] \left [
\mu_3^{expt}(\rho_4)- \mu_4^{expt}(\rho_4) \right ].
\end{equation}
  At the same time, an upper bound to the potential energy per particle 
  can be determined by 
\begin{equation}
v(\rho_4) \leq v^{UB}(\rho_4) \equiv e^{expt}(\rho_4) - t^{LB}(\rho_4)
\end{equation}
where $e^{expt}(\rho_4)$ is the $^4$He energy per particle. 
At the experimental saturation density, $\mu_4^{exp}(\rho_0)= -7.17$ K
 and $\mu_3^{exp}(\rho_0)=-2.78$ K, therefore
 $t_4^{LB}(\rho_0)=13.4 $ K and $v^{UB}(\rho_0)=-20.6$ K. 
These bounds are basically respected by all the theoretical calculations  
performed with the Aziz potential. Also the extraction of the kinetic
energy from the most recent inelastic neutron scattering data
is consistent with the kinetic energy lower-bound \cite{azuah97}. 

\section {The excitation spectrum of the $^3$He impurity}

\subsection {Variational calculation}  

Let us focus now on the variational description of the impurity 
 excited states. For pure $^4$He, a simple trial wave function for an
excited state characterized by a momentum $\vec q$ may be built in terms 
of Feynman phonons \cite{feynman54}
\begin{equation}
\Psi_{{\bf q}}(N_4)
 = 
 \rho_4({\bf q}) \Psi_0(N_4),
\end{equation}
where $\Psi_0(N_4)$ is the $^4$He ground state and 
\begin{equation}
\rho_4({\bf q})=\sum_{i=1,N_4} e^{i{\bf q} \cdot {\bf r}_i}
\end{equation}  
is the $^4$He density fluctuation operator. The excitation energy, 
$\omega (q)$,  of $\Psi_{{\bf q}}(N_4)$ is 
\begin{eqnarray}
\nonumber
\omega(q)&=&\frac {\langle \Psi_{{\bf q}}(N_4) \mid H - E_0 \mid \Psi_{{\bf q}}(N_4) \rangle}
{\langle \Psi_{{\bf q}}(N_4) \mid \Psi_{{\bf q}}(N_4) \rangle } 
  = 
\frac {\langle \Psi_0(N_4) \mid \rho_{{\bf q}}^{\dag} \rho_{{\bf q}} (H-E_0) \mid \Psi_0(N_4) 
\rangle }{\langle \Psi_{{\bf q}}(N_4) \mid \Psi_{{\bf q}}(N_4) \rangle }  
 \\  & + &
\frac {\langle  \Psi_0(N_4) \mid \rho_{{\bf q}}^{\dag} 
\left [ H -E_0 , \rho_{{\bf q }} 
\right ] \mid \Psi_0(N_4) \rangle }{\langle \Psi_{{\bf q}}(N_4) 
\mid \Psi_{{\bf q}}(N_4) \rangle}
 \, . 
\end{eqnarray}
The first term of the right hand side vanishes 
when $\Psi_0(N_4)$ is the exact ground state or contains  optimal two-body 
correlations. The second term leads to the well known Feynamn 
dispersion relation:
\begin{equation}
\omega (q)=\omega_F(q)=
\frac {\hbar^2 q^2 }{ 2m S(q)},
\end{equation}
where $S(q)$ is the static structure function defined in Eq.(\ref{eq:static}).
This relation gives the correct low momentum ($q \leq 0.4 \AA^{-1}$) 
 linear behavior of the phonon-roton spectrum, 
$\omega (q) = \hbar q v_s$, where $v_s$ is the speed of sound in 
liquid $^4$He ($v_s(\rho_0) \sim 238 m/s $), 
 provided that 
\begin{equation}
\lim_{q \rightarrow 0} S(q) = \frac {\hbar q }{2 m_4 v_s} .
\end{equation}

In an analogous way, the impurity excitated stated is 
given by 
\begin{equation} 
\Psi_{{\bf q}}(3;N_4)
 =\rho_3({\bf q}) \Psi_0(3;N_4),
\end{equation} 
where  $\rho_3({\bf q})=e^{ i{\bf q} \cdot {\bf r}_3}$ is the 
impurity excitation operator.
The  excitation energy is
\begin{equation}
\epsilon_0(q) = \frac {\langle \Psi_{{\bf q}}(3;N_4) \mid H(3;N_4)- E_0(3;N_4)
\mid \Psi_{{\bf q}}(3;N_4) \rangle }{\langle \Psi_{{\bf q}}(3;N_4) \mid 
\Psi_{{\bf q}}(3;N_4) \rangle }= \frac {\hbar q^2}{2 m_3},
\end{equation}
and one obtains a dispersion relation corresponding to the free particle, 
with $m_3^*=m_3$.
 
A better ansatz, which takes into account the backflow of $^4$He atoms 
around $^3$He, is provided by 
\begin{equation}
\Psi^{BF}_{{\bf q}}(3;N_4)= \rho_3({\bf q}) F_B({\bf q};3;N_4) \Psi_0(3;N_4)
\end{equation}
with 
\begin{equation}
F_B({\bf q};3;N_4) = \prod_{i=1,N_4} \exp {\left [i {\bf q}\cdot ({\bf r}_i - 
{\bf r}_3) \eta(r_{3i}) \right ]}.
\end{equation}
Backflow correlations do not change the binding energy of the impurity and
only affect the excitation spectrum, which, however, 
remains parabolic:
\begin{equation}
\epsilon_{BF}(q)=\frac {\hbar^2 q^2}{2 m_3}\left [1+a_1+a_2+a_3 \right ],
\end{equation}
where
\begin{equation}
a_1=\rho_4 \int d{\bf r} g^{(3,4)}(r) \left ( 2 \eta(r) + 
\frac {2}{3} r \eta'(r) \right ),
\end{equation}
\begin{equation}
a_2= \frac {m_3}{\mu_{red}} \rho_4 \int d {\bf r} g^{(3,4)}(r) \left (
\eta(r)^2 + \frac {1}{3} \left [ r^2 (\eta'(r))^2 + 2 \eta(r)
r \eta'(r) \right ] \right ),
\end{equation}
\begin{equation}
a_3 = \rho_4^2 \int d {\bf r}_{31} d {\bf r}_{32} 
g^{(3,1,2)} \left ( \eta_{31} 
\eta_{32} + \frac {1}{3} \left [ r_{31} \eta_{31}' \eta_{32}'
r_{32}(\hat {\bf r}_{31} \cdot \hat {\bf r}_{32} )^2 + 2 \eta_{31} \eta_{32}' 
r_{32} \right ] \right ),
\end{equation} 
where $g^{(3,1,2)}$ is a three body distribution function. 
A good choice for the  variational function $\eta(r_{3i})$ has been 
found to be
\begin{equation}
\eta(r)= A_0 \exp{\left ( - \left [ \frac {(r-r_0)}{\omega_0}\right ]^2
 \right )},
\end{equation}
and the parameters $r_0$, $\omega_o$ and $A_0$ are varied to find the minimum.
With this variational ansatz, one gets a $q$-independent effective mass of  
$m_3^*/m_3\sim 1.7$. A minimization through a full functional 
variation with respect to $\eta(r_{31})$ \cite{owen} does  not 
change this result.

\subsection{Correlated perturbative approach}

A systematic way to incorporate effects more complicated than the 
two--body back-flow is provided by the
Correlated Basis Function perturbation theory. In such an approach,  
 CBF is used to build   richer excitations on top of 
$\Psi_{{\bf q}}$ by allowing the total momentum to be shared between 
the impurity and the phonons in the medium. The underlying idea 
consists in incorporating  the non perturbative correlations
directly into the basis functions and then developing a perturbative 
expansion in this basis. Usually, since the correlated states 
are a good approximation to the eigenstates of the hamiltonian, the 
first low orders of the perturbation theory
are sufficient for getting reliable and realistic results.
 
In the case of the impurity, the correlated basis is classified according
 to the number of Feynman phonons in each intermediate state,
\begin{equation}
\Psi_{{\bf q};{\bf q}_1..{\bf q}_n}(3;N_4)
 =\rho_3({\bf q}-{\bf q}_1-..-{\bf q}_n) 
 \rho_4({\bf q}_1)..\rho_4({\bf q}_n) \Psi_0(3;N_4).
\label{eq:nphonon}
\end{equation}
The multiphonon states will mix in the perturbative process with the 
unperturbed state $\Psi_{\bf q}= \rho_3({\bf q}) \Psi_0$.
The states (\ref{eq:nphonon}) are not orthogonal among each other and 
either they must be properly normalized following the procedure 
described in Ref.(34) or the perturbative expansion 
must be carried on in a non orthogonal basis \cite{morse}. 

The results presented in this work have been obtained including 
one phonon (OP) and two independent phonon (TIP) intermediate 
states and all the perturbative diagrams corresponding to successive 
rescatterings of the one phonon states (OPR). 
While the correlation factors are intended to care for the short 
range modifications of the ground state wave function due to the 
strongly repulsive interatomic potential, the basic physical effect 
induced by the perturbative corrections may be traced back to 
different types of backflow around both the impurity and the 
$^4$He atoms.  


In order to construct the CBF perturbative series,
 the unperturbed and the interaction Hamiltonians are defined
via  their matrix elements:
\begin{equation}
H_{0,ij}= \delta_{ij} \frac{ \langle \Psi_i \mid H \mid \Psi_j \rangle}
{\langle \Psi_i \mid \Psi_i \rangle }= \delta_{ij} E_i^v,
\end{equation}
\begin{equation}
H_{I,ij}=(1-\delta_{ij}) \frac {\langle \Psi_i \mid H - E_q \mid \Psi_j \rangle}
{(\langle \Psi_i \mid \Psi_i \rangle \langle \Psi_j \mid \Psi_j \rangle)^{1/2}}=
(1 - \delta_{ij}) (H_{ij} - E_q N_{ij}),
\end{equation}
where $E_q=E_0(3;N_4)+\epsilon_0(q)+ \delta\epsilon(q)$ is the eigenvalue
of $H$ for the state with momentum ${\bf q}$.  
The diagonal matrix elements of $H_I$ are zero by construction 
and, therefore, there are no first order perturbative corrections.

The Brillouin-Wigner series for the perturbative correction 
$\Delta E_q$ to $E_q^v$,  
appropriate for non-orthogonal states, is given by:
\begin{equation}
\Delta E_q = \sum_{j\neq q } \frac {(H_{qj} - E_p N_{qj})(H_{jq} - E_p N_{jq})}
{E_q - E_j^v} + ..... ,
\label{eq:BW}
\end{equation}
where $E_q$ is the final energy and the overlap matrix elements, $N_{qj} =
\langle \Psi_q \mid \Psi_j \rangle$, take care of the non-orthogonalization. 
The series is first expanded around the correction to the medium 
energy, $\Delta E_0 = E_4 -E_4^v$,  and then 
resummed in such a way to cancel all the terms diverging  
in the $N_4\rightarrow \infty$ limit and originating  from 
both the cluster expansion of the matrix elements and the 
perturbative expansion itself. 
This procedure has been devised for the ground-state energy of an infinite
Fermi system \cite{fantoni84} and then generalized to the case of the impurity 
\cite{fabro86}. 

The difficulties related to the divergencies  
can be avoided by working in the orthogonalized scheme, in fact 
the $n$-phonon states may be Schmidt-orthogonalized to 
states with a lower number of phonons. For instance, the orthogonalized 
OP state reads as:
\begin{equation}
|{\bf q};{\bf q}_1\rangle = \frac {
|\Psi_{{\bf q};{\bf q}_1}\rangle- 
|\Psi_{{\bf q}}\rangle
\langle \Psi_{{\bf q}}|\Psi_{{\bf q};{\bf q}_1}\rangle}{
\langle \Psi_{{\bf q};{\bf q}_1}|\Psi_{{\bf q};{\bf q}_1}\rangle^{1/2}}
 \, , 
\label{eq:OP_0}
\end{equation}
while the two-phonon state, $\Psi_{{\bf q};{\bf q}_1{\bf q}_2}$, may 
be orthogonalized in a similar way to 
$\Psi_{{\bf q}}$, $\Psi_{{\bf q};{\bf q}_1+{\bf q}_2}$ and  
$\Psi_{{\bf q};{\bf q}_{1,2}}$. The orthogonalizaton makes the 
convergence of the series faster as the non orthogonalized 
states have large mutual overlaps 

The non diagonal matrix elements  of the hamiltonian, $H$, 
can be easily evaluated by 
assuming that the two- and three-body correlations are solutions 
of the corresponding Euler equations. Within this assumption, 
 matrix elements involving $n$ phonons may be expressed in 
terms of ($n+1$)--body structure functions. As an example, 
the matrix element between the OP state and the one without phonons, 
$|{\bf q}\rangle$, is given by 
\begin{equation}
\langle{\bf q}|H |{\bf q};{\bf q}_1\rangle = 
-{[N_4 S(q_1)]^{-1/2}}
\frac {\hbar^2}{2 m_3} {\bf q}\cdot {\bf q}_1 S^{(3)}(q_1)
 \, , 
\label{eq:OP_1}
\end{equation}
where $S(q)$ and $S^{(3)}(q)$ are the two--body structure functions of eqs.
(\ref{eq:static}) and (\ref{eq:staticI}).

The diagonal matrix elements have the particularly simple form:
\begin{equation}
\langle {\bf q};{\bf q}_1..{\bf q}_n|H|
{\bf q};{\bf q}_1..{\bf q}_n\rangle  = 
E_0^v+\epsilon_0(q)+\sum_{i=1,n}w_F(q_i)
\label{eq:diag_2}
\end{equation}
with $E_0^v=\langle \Psi_0|H|\Psi_0\rangle/\langle \Psi_0|\Psi_0\rangle$. 

 Finally, the impurity excitation energy is 
$\epsilon(q)=\epsilon_0(q)+\Delta \epsilon(q)$, where, within the 
truncation we have used,  
\begin{equation}
\Delta \epsilon (q)\sim\Delta \epsilon_{OP}(q)+\Delta \epsilon_{TIP}(q)
 +\Delta\epsilon_{OPR}(q)
 \, . 
\label{eq:Delta}
\end{equation}
The different terms in (\ref{eq:Delta}) represent contributions 
from the corresponding intermediate states.

The OP and TIP contributions to $\Delta \epsilon(q)$ are
shown in Figure 5 of Ref.(5). We stress that the CBF based perturbative 
approach requires the evaluation of  two levels of diagrams, 
the perturbative and the cluster ones. 
We use Brillouin-Wigner perturbation theory, so 
the correction itself depends on $\epsilon(q)$ and the series must be 
summed self-consistently. For instance, the OP contribution 
is solution of 
\begin{eqnarray}
\nonumber
\Delta \epsilon_{OP}(q)&=&\sum_{{\bf q}_1} \frac { \vert 
\langle {\bf q}|H-E_0-\epsilon(q)|{\bf q};{\bf q}_1\rangle \vert ^2
}{
\epsilon(q)-\epsilon_0(|{\bf q}-{\bf q}_1|)-w_F(q_1)}
 \\  & = &
\frac {\Omega}{(2\pi)^3}
\left(\frac {\hbar^2}{2m_3}\right)^2
\int d^3q_1
\frac {1}{N_4S(q_1)}
\frac {[S^{(3)}(q_1){\bf q}\cdot{\bf q}_1]^2}{
\epsilon(q)-\epsilon_0(|{\bf q}-{\bf q}_1|)-w_F(q_1)}
 \, , 
\label{eq:OP_3}
\end{eqnarray}
and the OP effective mass, at $q=0$, is 
\begin{equation}
\frac {m_3^* }{m_3} = \left 
[1 - \frac {1}{4 \pi^2 \rho_4} \frac {\hbar^2}{2 m_3}
\frac {2}{3} \int dq \frac 
{q^2 S^{(3)}(q)^2}{\frac {S(q) \hbar^2}{2 m_3} + \frac {\hbar^2}
{2 m_4 }} \right ]^{-1} .
\end{equation}

The spectrum obtained by taking only OP states  is very close to the 
LP one with an effective mass similar to that given by 
the variational calculation with backflow correlations.
Actually, in several papers it has been 
pointed out that second order perturbative expansion with 
OP states introduces two--body backflow correlations into the wave 
function \cite{fabro86,feenberg69}. We find 
$m^*_3(OP)=1.8~m_3$, in good agreement with an analogous CBF treatment by 
Saarela \cite{saarela90}  ($m_3^*\sim 1.9~m_3$)
and with the backflow variational calculations\cite{fabro86}. 

The matrix elements involving TIP states 
(whose lenghty expressions are given in Ref.(5)), 
involve, as already anticipated,  the two-- and three--body 
structure functions, i.e. the Fourier transforms of the two-- 
and three--body distribution functions, 
$g^{(2)}(r_{12})$ and $g^{(3)}({\bf r}_1,{\bf r}_2,{\bf r}_3)$. 
$g^{(3)}$ may be evaluated within several approximations, the most 
common of which are the convolution (CA) and the superposition (KSA) 
ones \cite{feenberg}. 
The CA correctly accounts for the sequential relation between 
$g^{(3)}$ and $g^{(2)}$ and factorizes in momentum space, 
$S^{(3)}_{CA}({\bf q}_1,{\bf q}_2,{\bf q}_3)=S(q_1)S(q_2)S(q_3)$. 
The SA factorizes in $r$-space, 
$g^{(3)}_{KSA}({\bf r}_1,{\bf r}_2,{\bf r}_3)=
g^{(2)}(r_{12})g^{(2)}(r_{13})g^{(2)}(r_{23})$, and 
adequately describes the short range region. 
The momentum space factorization property makes the 
CA more suited to our perturbative study.

The two approximations give $m_3^*(CA)=1.6~m_3$ and $m_3^*(KSA)=2.1~m_3$ 
at $q=0$, with OP and TIP states. 
A calculation including a four--body correction (the Abe term \cite{abe})
 in the three-body distribution function and the OPR contribution provides
$m_3^*= 2.2~m_3$ at  saturation density \cite{fabro86}. 
 Moreover, a key ingredient for a correct behavior of $\epsilon (q)$ 
in the large $q$ sector is a good description of the $^4$He roton. 
This requires the use of the superposition
approximation that, on the other hand, is not correct in the 
$^4$He phonon region. The inclusion of backflow correlations in the 
correlated intermediate states gave an overall agreement with the 
experimental $^4$He spectrum \cite{manou86} but largely increased  
both the difficulty of the evaluation of the matrix elements and the 
uncertainty related to the use of high order distribution functions. 
 In the impurity case a good compromise (termed CA1), 
which does not require a big computational effort, 
was found by using the CA and the experimental values of the $^4$He spectrum, 
 $\omega_{expt}(q)$,  in the energy denominators. 

In Figure 2, we show the  impurity spectrum in CA1, 
together with the experimental $^3$He and $^4$He curves. 
The OPR terms are included and the LP and MLP fits to 
$\epsilon_{expt}(q)$ are shown. Since the 
branch of the dynamical response due to the excitations of the low 
concentration $^3$He component in the Helium mixtures overlaps 
the collective $^4$He excitation at $q>1.7~\AA^{-1}$ 
\cite{Fak90}, $\epsilon_{expt}(q)$ is not known in 
that region.  
\begin{figure}
\begin{center}
\epsfysize=2.5in \epsfbox{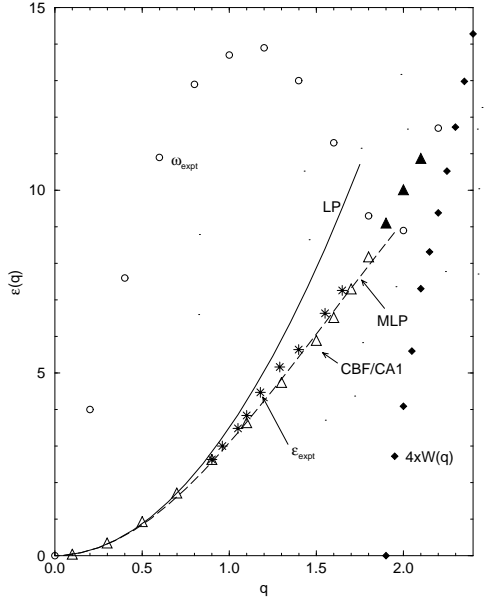}
\end{center}
\caption{CBF/CA1 (triangles), LP and MLP $^3$He single particle energies (in K).
Stars and circles are the impurity and $^4$He experimental data, respectively. 
Black triangles are extrapolated CBF/CA1 values. 
Black diamonds give the imaginary part
of the impurity self-energy (in K). Momenta in $\AA^{-1}$}
\end{figure}
 
The CBF spectrum is very close to $\epsilon_{expt}(q)$ 
up to its merging into the $^4$He dispersion relation. 
For the $\gamma$  parameter in the MLP parametrization, the theory gives 
$\gamma (CBF)\sim 0.19~\AA^2$ and $m_3^*(CBF)=2.1~m_3$. If the spectrum
 is parametrized in terms of a momentum dependent effective mass, 
$\epsilon(q)=\hbar^2 q^2/2m_3^*(q)$, then we find 
$m^*_3(q=1.7~\AA^{-1})=3.2~m_3$, with an increase of $\sim 50\%$ 
respect to the $q=0$ value. 
 
Beyond $q\sim 1.9~\AA^{-1}$, the impurity quasiparticle 
is no longer an excitation with a well defined 
energy and it can decay into $^4$He excitations and acquire a 
finite lifetime, $\tau$. A finite $\tau$--value reflects a non zero 
imaginary part of the $^3$He complex  
on-shell self-energy, $W(q)=\Im \Sigma (q,\epsilon(q))$. 
 $W(q)$, amplified by a factor 4 and  computed with only
 OP intermediate states, is shown in Figure 2. Our 
$W(q)$ is close to the 
one found in Refs.\cite{saarela90,kro98}. A numerical extrapolation
 of the computed $\epsilon_{CBF}(q)$ into the roton region does not show any 
evidences of a $^3$He roton-like structure. 

Shadow wave functions have been used also for studying the excited states of
the impurity. In  this  case, an excited state
of momentum ${\bf q}$ is constructed by associating the momentum to the 
impurity shadow variables, 
\begin{equation}
\Psi_{\bf q}^3 (R) = \int dS F(R,S) \tilde \delta_{{\bf q}}. 
\end{equation}
Similarly, a $^4$He excited states is built as:
\begin{equation}
\Psi_{\bf q}^B(R) = \int dS F(R,S) \tilde \sigma_{{\bf q}}.
\end{equation}
The momentum is carried on by  the shadow variables with
\begin{equation}
\tilde \delta_{{\bf q}}= e^{i {\bf q}\cdot \tilde {\bf s}_3} , ~~~~~~~~~~~~
\tilde \sigma_{{\bf q}}= \sum_{j=1}^{N_4} e^{i {\bf q}\cdot \tilde {\bf s}_j}
\end{equation}
where $\tilde {\bf s}_{j,3}$ are shadow variables  modified by an
explicit back-flow correlation
\begin{equation}
\tilde {\bf s}_j = {\bf s}_j + \sum_{l\neq j} 
({\bf s}_j -{\bf s}_l) \lambda(s_{jl}).
\end{equation}
The two shadow wave functions are eigenstates of the momentum operator, 
both with momentum ${\bf q}$, and are therefore orthogonal to the ground state.
 However, they are not orthogonal between them. 
They must be orthogonalized  and a $2 \times 2$ Hamiltonian must be 
diagonalized for each momentum in order to properly consider the 
collective and the single-particle impurity branches.  
This procedure has been followed by Galli et al.\cite{reatto99} 
by a Monte Carlo algorithm and the final spectrum is compatible
with a MLP spectrum with parameters $m_3^*=2.06~m_3$ and  
$\gamma = 0.0314~\AA^2$.
The spectrum is close to the LP form and $m_3^*=1.74~m_3$ when 
there is no back-flow term in the shadow variables 
($\lambda(r_{ij})\equiv 0$). 

Other methods have confronted themselves with the impurity problem 
in the many--body arena. We first mention diffusion Monte Carlo, who 
has recently given an effective mass of  $m_3^*/m_3=2.2$ \cite{boro99}
 at saturation density, at zero momentum. 
Another one is based on a dynamic theory that allows
for time dependent correlations, whose equations of motion are
determined through a minimum  action principle. 
This theory has been succesfully applied
to study the excitation spectrum of pure $^4$He \cite{saarela86}. 
The method is very closely  related to 
CBF and, in a simplifying hypotesis known as "uniform limit
approximation",  one recovers the second order expression in the OP space.
The resulting spectrum is well adjusted to the MLP form with
$m_3^*=2.09~m_3$ and $\gamma=0.114~\AA^2$ at saturation density 
\cite{kro98}.    

\section {Conclusions}
 
We have shown in this contribution that a variational theory employing 
 correlated wave functions is able to provide a good description of the 
ground state of one $^3$He impurity in liquid $^4$He. 
In addition, a perturbative expansion in a correlated basis may 
give a quantitative picture of the impurity 
excitation spectrum, provided  the basis considers 
correlated states with two independent phonons and one phonon 
rescattering diagrams, which play a non marginal role at large 
momenta. The agreement with other many--body methodologies, as shadow 
wave functions, diffusion Monte Carlo and time dependent correlations, 
is quite good. This gives confidence in the possibility of extending 
the CBF theory to other, less studied, aspects of the physics of liquid 
Helium. To stay in the field of the mixtures,  first CBF analyses of the 
inelastic neutron scattering cross sections,  
both at low  and high momentum transfers \cite{boronat93,vichi}, 
have revealed an encouraging  semiquantitative agreement with the 
available experimental results \cite{Fak90,sokol}. This topic needs 
to be more carefully investigated, also within a close interaction with 
the experimental teams working, or willing to work, in this subject. 

Finally, to conclude this presentation, we rest our case.
 

\section*{Acknowledgments}

The authors  have profited from  fruitful discussions and collaborations
with J. Boronat, C. E. Campbell, S. Fantoni, E. Krotscheck 
 and F. Mazzanti. This work has been supported by DGICYT (Spain)
Grant No. PB95-1249, the agreement CICYT (Spain)-INFN (Italy) and 
the program SGR98-11 from Generalitat de Catalunya.

\eject


\begin{thebibliography}{99}

\bibitem{booktwo} {\it Excitations in Two-Dimensional and Three-Dimensional 
Quantum Fluids} , Vol. 257 of NATO {\it Advanced Study Institute, Series B:
Physics}, A.F.G. Wyatt and H.J. Lauter, eds.( Plenum, New York, 1991). 
\bibitem{hallock99} W. Teizer, R. B. Hallock, E. Dujardin, and T. W. Ebbesen,
 \Journal{\PRL}{82}{5305}{1999}  
\bibitem{vilesov99} S. Grebenev, J. P. Toennies and A. Vilesov, 
\Journal{\SCI}{279}{2083}{1998}
\bibitem{book1} {\em Microscopic Quantum Many-Body Theories and Their Applications},
 J. Navarro and A. Polls eds. , Lecture Notes in Physics, Vol. 510, (Springer-Verlag, 
Heidelberg, 1998)
\bibitem{fabro86} A. Fabrocini, S. Fantoni, S. Rosati and A. Polls,
\Journal{\PRB}{33}{6057}{1986}
\bibitem{fabro98} A. Fabrocini and A. Polls, \Journal{\PRB}{58}{5209}{1998}
\bibitem{reatto99} D. E. Galli, G.L. Masserini and L. Reatto, \Journal{\PRB}{60}
{3476}{1999}
\bibitem{kro98} E. Krotscheck, J. Paaso, M. Saarela, K. Sch\o okhuber and R. Zillich
\Journal{\PRB}{58}{12282}{1998}  
\bibitem{boro99} J. Boronat and J. Casulleras, \Journal{\PRB}{59}{8844}{1999}
\bibitem{Fak90}  B. F{\aa}k, K. Guckelsberger, M. Korfer, R. Scherm and 
A. J. Dianoux,\Journal{\PRB}{41}{8732}{1990}.
\bibitem{boronat89} J. Boronat, A. Fabrocini and A. Polls,\Journal{\PRB}
{39}{2700}{1989}
\bibitem{wilks67} J. Wilks in {\em The properties of Liquid and Solid Helium}(Clarendon
Press, Oxford,1967)
\bibitem{ebner70} C. Ebner and D. O. Edwards,\Journal{\PHYS}{2}{77}{1970}. 
\bibitem{Landau48} L. D. Landau  and I. M. Khalatnikov, \Journal{\Soviet}
 {2}{637}{1948}
\bibitem{kro198} E. Krotscheck, M. Saarela, K. Sch\"orkhuber, and R. Zillich, 
\Journal{\PRL}{80}{4709}{1998}
\bibitem{yoro93} S. Yorozu, H. Fukuyama, and H. Ishimoto, \Journal{\PRB}{48}{9660}
{1993}
\bibitem{Aziz} R. A. Aziz, V. P. S. Nain, J. S. Carley, W. L. Taylor, and G.T. 
McConville, \Journal{\CHEM}{70}{4330}{1979}
\bibitem{Aziz1} R. A. Aziz, F. R. W. McCourt, and C. C. K. Wong, \Journal{\MOL}
{61}{1487}{1987}
\bibitem{feenberg} E. Feenberg, {\em Theory of Quantum Liquids},
 Academic, New York, 1969.  
\bibitem{moroni95} S. Moroni, S. Fantoni, and G. Senatore, \Journal{\PRB}{52}
{13547}{1995}
\bibitem{fabrocini84} A. Fabrocini and A. Polls, \Journal{\PRB}{30}{1200}{1984}
\bibitem{schmidt} K. Schmidt, M.H. Kalos, M.A. Lee, and G.V. Chester, \Journal
{\PRL}{45}{573}{1980}
\bibitem{usmani} Q.N. Usmani, S. Fantoni, and V.R. Pandharipande, \Journal{\PRB}
{26}{6123}{1982}
\bibitem{saarela93} M. Saarela and E. Krotscheck, \Journal{\LOW}{90}{415}{1993}
\bibitem{saarel193} E. Krotscheck and M. Saarela, \Journal{\PHYS}{232}{1}{1993}
\bibitem{fabro199} A. Fabrocini and S. Fantoni in {\em Advances in Quantum Many-Body 
Theories}, Vol. 2, R.F. Bishop and N.R. Wilet eds.,(World Scientific, Singapore,1998) 
\bibitem{polls93} J. Boronat, A. Fabrocini, and A. Polls, \Journal{\LOW}{74}{347}
{1989}
\bibitem{baym66} G. Baym, \Journal{\PRL}{17}{952}{1966}
\bibitem{vitiello} S. A. Vitiello, K. Runge, and M. H. Kalos, \Journal{\PRL}
{60}{1970}{1988}
\bibitem{moroni} S. Moroni, D. E. Galli, S. Fantoni, and L. Reatto, \Journal{\PRB}
{58}{909}{1998}
\bibitem{azuah97} R.T. Azuah, W.G. Stirling, H.R. Glyde, M. Bonisegni, 
P.E. Sokol, and S.M. Bennington, \Journal{\PRB}{56}{14620}{1997}
\bibitem{feynman54} R. P. Feynman, \Journal{\PR}{94}{262}{1954} 
\bibitem{owen} J. C. Owen, \Journal{\PRB}{23}{5815}{1981}
\bibitem{fantoni88} S. Fantoni and V. R. Pandharipande, \Journal{\PRC}{37}{1697}{1988}
\bibitem{morse} P. M. Morse and H. Feshbach, 
{\it Methods of Theoretical Physics},
(McGraw-Hill, New York), 1953.
\bibitem{fantoni84} S. Fantoni, \Journal{\PRB}{29}{2544}{1984}
\bibitem{feenberg69} T. B. Davison and E. Feenberg, \Journal{\PR}{178}{306}{1969}
\bibitem{saarela90} M. Saarela, {\it Recent Progress in Many Body Theories}, ed. Y. Avishai
(Plenum, New York, 1990), Vol. 3, p. 337.
\bibitem{abe} R. Abe, \Journal{\PRO}{21}{421}{1959}
\bibitem{manou86} E. Manousakis and V. R. Pandharipande, \Journal{\PRB}{33}{150}{1986}
\bibitem{saarela86} M. Saarela, \Journal{\PRB}{33}{4596}{1986}
\bibitem{boronat93} J. Boronat, F. Dalfovo, F. Mazzanti and A. Polls, 
\Journal{\PRB}{48} {7409} {1993}
\bibitem{vichi} A. Fabrocini, L. Vichi, F. Mazzanti, and A. Polls, \Journal{\PRB}{54}{10035}
{1996}
\bibitem{sokol} Y. Wang and P. E. Sokol, \Journal{\PRL}{72}{1040}{1994}


\end{thebibliography}
\end{document}